\documentclass[10pt,  twocolumn, twoside, letterpaper, conference, final]{IEEEtran}
\IEEEoverridecommandlockouts
\usepackage{cite}
\usepackage{algorithmicx}
\usepackage[ruled]{algorithm}
\usepackage{algpseudocode}
\usepackage{mathtools}
\usepackage{dsfont}

\usepackage{amsmath}
\usepackage{amsfonts}
\usepackage{amssymb}

\usepackage{graphicx}
\usepackage{subcaption}
\usepackage{multirow}
\usepackage{textcomp}
\def\BibTeX{{\rm B\kern-.05em{\sc i\kern-.025em b}\kern-.08em
    T\kern-.1667em\lower.7ex\hbox{E}\kern-.125emX}}
\usepackage{color}
\usepackage[keeplastbox]{flushend}
\usepackage{enumerate}
\usepackage{float}
\usepackage{caption}
\usepackage{mathrsfs}
\usepackage{mathtools}

\usepackage{mathtools}

\usepackage[english]{babel}
 
\newtheorem{theorem}{Theorem}

\makeatletter
\newcommand\fs@spaceruled{\def\@fs@cfont{\bfseries}\let\@fs@capt\floatc@ruled
  \def\@fs@pre{\vspace{0.5\baselineskip}\hrule height.8pt depth0pt \kern2pt}%
  \def\@fs@post{\kern0pt\hrule\relax}%
  \def\@fs@mid{\kern0pt\hrule\kern0pt}%
  \let\@fs@iftopcapt\iftrue}
\makeatother
\usepackage{caption}

\title{Physical Layer Security: Authentication, Integrity and Confidentiality}
\author{\IEEEauthorblockN{Mahdi Shakiba-Herfeh}
\IEEEauthorblockA{\textit{ETIS, Universit\'e Paris Seine, ENSEA } \\
\textit{Universit\'e Cergy-Pontoise,  CNRS}\\
 Cergy-Pontoise, France \\
mahdi.shakiba-herfeh@ensea.fr}
\and
\IEEEauthorblockN{Arsenia Chorti}
\IEEEauthorblockA{\textit{ETIS, Universit\'e Paris Seine, ENSEA } \\
\textit{Universit\'e Cergy-Pontoise,  CNRS}\\
 Cergy-Pontoise, France \\
arsenia.chorti@ensea.fr}
\and
\IEEEauthorblockN{H. Vincent Poor}
\IEEEauthorblockA{\textit{Department of Electrical Engineering} \\
\textit{Princeton University}\\
 Princeton, NJ\\
poor@princeton.edu}
}

\begin{document}

\maketitle
\begin{abstract}
The goal of physical layer security (PLS) is to make use of the properties of the physical layer -- including the wireless communication medium and / or the transceiver hardware -- to enable critical aspects of secure communications. In particular, PLS can be employed to provide i) node authentication, ii) message authentication, and, iii) message confidentiality. Unlike the corresponding classical cryptographic approaches which are all based on computational security, PLS's added strength is that it is based on information theoretic security, in which no limitation with respect to the opponent's computational power is assumed and is therefore inherently quantum resistant. In this survey, we review the aforementioned fundamental aspects of PLS, starting with node authentication, moving to the information theoretic characterization of message integrity, and finally, discussing message confidentiality both in the secret key generation from shared randomness and from the wiretap channel point of view. The aim of this review is to provide a comprehensive road-map on important relevant results by the authors and other contributors and discuss open issues on the applicability of PLS in sixth generation systems.
\end{abstract}

\begin{IEEEkeywords}
Physical layer security, physical unclonable function, RF fingerprinting, node authentication, message integrity, secrecy encoder, shared randomness, key generation, confidentiality.
\end{IEEEkeywords}

\section{Introduction}
The increasing deployment of wireless systems poses security challenges in next generation dynamic and decentralized networks, consisting of low cost and complexity devices. Over the last two decades alternative / complementary means to secure data exchange in wireless settings have been investigated in the framework of physical layer security (PLS), addressing jointly the issues of reliability and secrecy.
PLS takes advantage of the inherent randomness of wireless communication channels and / or the unclonability of hardware fabrication processes, to harvest entropy and deliver authentication, confidentiality, message integrity and privacy in demanding scenarios. In this chapter, we revisit all aforementioned aspects from an information theoretic security perspective.

PLS relies on information theoretic proofs of (weak or strong) perfect secrecy, a notion first introduced by Shannon in 1949 \cite{Shannon49}. As such, PLS systems cannot be ``broken'' irrespective of the adversarial computational power, i.e., the proofs do not rely on any assumptions regarding the hardness of particular families of algebraic problems. There are some fundamental differences between information theoretic security and  classical cryptography  based security. In the following, we illustrate some of the pros and cons for each.

\begin{figure}[t]
\centering
\includegraphics[width=0.48\textwidth]{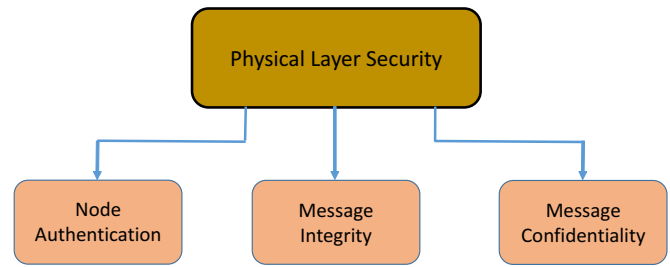}
\caption{The three main operations of PLS.}
\label{diagram}
\end{figure}

\textit{Classical cryptography based security}:
standard cryptosystems as those in employed in the fifth generation (5G) security protocols have notable strengths:
\begin{itemize}
   \item There are no known feasible attacks on symmetric key cryptosystems such as the advanced encryption system (AES) or elliptic curve Diffie Hellman (ECDH) asymmetric key encryption, and hence they are trustworthy in any conceivable scenario as they are thought of achieving semantic security;
   \item Only a few assumptions are made about the messages to be encrypted or the trusted third parties in authentication protocols (e.g., regarding the trustworthiness of certificate authorities);
   \item These systems have been widely employed and tested over decades, the technology is mature, ready-to-use and nowadays inexpensive.
\end{itemize}

However, crypto based security has indeed certain disadvantages, some of which are pointed out below:
\begin{itemize}
   \item Generally the semantic security proofs of traditional crypto systems are built around unproven assumptions about the hardness of certain ``one-way'' functions. As a result, some of these schemes, notably in the realm of asymmetric key encryption, are considered vulnerable to quantum attacks;
   \item Standard crypto is typically employed in upper layers of the OSI protocol stack, assuming that the PHY connection has already been established. As a result, they are inherently ``inflexible'' with respect to wireless connectivity issues and will fail in attacks at the physical layer, e.g., jamming attacks on the control plane;
   \item State-of-the art key distribution schemes for wireless networks based on the classic cryptography model require a trusted third party and are typically computationally intensive. Therefore, their application in machine-to-machine or low latency applications can be challenging [reference urllc 3 gpp security];
   \item These security approaches are not tailored to the wireless communication properties and are typically not lightweight. With respect to the latter aspect, as an example, the level of sophistication of Google's take on a lightweight implementation of AES is rather a proof of the difficulty in rendering these schemes lightweight, rather than the opposite.
\end{itemize}

\textit{Information theoretic security}:
Notable advantages of PLS based security are as follows:
\begin{itemize}
   \item No computational limitations are placed on the opponent, PLS schemes that are properly implemented are quantum secure;
   \item The achievable secrecy rate is a function of the channel quality and the block length of the secrecy encoders and as a result the security is naturally tied to the communication properties;
   \item Unlike ``distributing'' keys, PLS can be used to generate on-the-fly secret keys, exploiting channel estimation operations that are customarily performed to establish the PHY connection.
   \item PLS implementations can be lightweight and related schemes can be advantageous in Internet of things (IoT) or low latency constrained scenarios.
\end{itemize} 

Also the disadvantages of this class of security are as follows
\begin{itemize}
   \item Some PLS schemes are based on stringent assumptions about the adversarial channel quality, e.g., the wiretap channel scenario, that are impractical in the general case;
   \item PLS technologies have not been tested ``in the field'' and it is therefore expected that there will be erroneous implementations before reaching a satisfactory level of maturity;
   \item The performance bounds of the related encoders have not been characterized in the finite blocklength regime, so the achievable rates back-off from the information theoretic infinite blocklenth capacity is yet unknown.
\end{itemize}

Despite these issues, PLS is currently studied as a possible second layer of security for particular use cases, e.g., when implementation issues in the 5G security protocols have identifiable shortcomings such as  vulnerabilities to false base station attacks \cite{3GPPFalseBase}. Notably, it is  explicitly mentioned as a sixth generation (6G) enabling technology in the first white paper on 6G: ``The strongest security protection may be achieved at the physical layer''. In this work, we review how it is possible to move some of the security core functions down to the physical layer, exploiting both the communication radio channel and the hardware as unique entropy sources.

We consider three important security operations: node authentication, message integrity and message confidentiality  as depicted in Fig. \ref{diagram}. In node authentication, the goal is for nodes to identify uniquely the other side of the communication. With respect to message integrity, the goal is to be able to identify tampering attacks on the exchanged messaged, i.e., verify the integrity of the received information in the presence of active attackers. Finally, in message confidentiality, users want to ``hide'' the content of their transmissions from a passive opponent (eavesdropper).

The rest of the paper is organized as follows. In Section \ref{NodeAuth} three different PLS methods of node authentication are reviewed:  i) physical unclonable functions (PUFs), ii) biometric based authentication, and iii) RF fingerprinting. Next, in Section \ref{MssgIntg}, the information theoretic bounds on the achievable rates when message integrity is required are reviewed, both for noiseless and noisy transmission channels. Furthermore, in Section \ref{MssgConf} we consider message confidentiality. Two alternative approaches to achieve message confidentiality are reviewed: i) keyless secrecy encoding in wiretap channels and, ii) channel based secret key generation (SKG), used in conjunction with symmetric encryption in hybrid schemes.

\section{Node authentication}\label{NodeAuth}
In all communication networks, users utilize authentication protocols to prove their identity. In standard crypto protocols, asymmetric key encryption is typically used in the authentication phase. However, the standard cryptographic schemes in the realm of public key encryption (PKE), are computationally intensive, incurring considerable overhead and can rapidly drain the battery of energy constrained devices \cite{Mukherjee15,Yener15}. Additionally, traditional public key generation schemes are not \textit{quantum secure} -- in that when sufficiently capable quantum computers will be available they will be able to break current known public key encryption schemes -- unless the key sizes increase to impractical lengths. 

\begin{figure*}[t]
\setlength{\unitlength}{0.17in} 
\centering 
\begin{picture}(27.5,7.5)
\put(0,3){\framebox(4,3)}
\put(-1,3.5){\line(1,0){7}}
\put(-1,5.5){\line(1,0){7}}
\put(6,3){\framebox(4,3)}
\put(6,3.5){\line(2,1){4}}
\put(6,5.5){\line(2,-1){4}}
\put(10,3.5){\line(1,0){2}}
\put(10,5.5){\line(1,0){2}}
\put(12,3){\framebox(4,3)}
\put(12,3.5){\line(2,1){4}}
\put(12,5.5){\line(2,-1){4}}
\put(16,3.5){\line(1,0){1}}
\put(16,5.5){\line(1,0){1}}
\put(17.5,4.5){$\dots$}
\put(19,3.5){\line(1,0){6}}
\put(19,5.5){\line(1,0){6}}
\put(20,3){\framebox(4,3)}
\put(25,2){\framebox(4,5)}
\put(29,4.5){\line(1,0){1}}
\put(30.3,4.2){Arbiter output}
\put(-5,2){Challenge: \hspace{1.1 cm}  $0$ \hspace{2.2 cm} $1$ \hspace{2.2 cm} $1$ \hspace{3.1cm} $0$ }
\end{picture}
\caption{Arbiter PUF}
\label{fig:arbiter_puf}
\end{figure*}
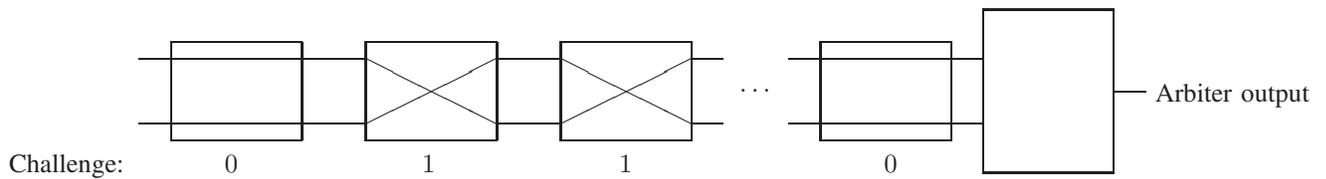

As a result, in 6G, PLS based authentication arises as a possible alternative. PLS authentication protocols usually consist of two stages, namely an enrollment stage and a release (authentication) stage. The enrollment stage occurs off-line. In this stage, unique characteristics of the node or user to be authenticated are measured. Hashed versions of these measurements along with related helper (side) information are stored at the verifier side in a database. In the release stage, new measurements are taken and sent to the verifier; the latter uses the helper information to regenerate the hash of the initial measurement, in which case the authentication is successful.  The role of the helper information is critical as it allows to correct for discrepancies between different measurements due to noise (in any actual system,  deriving the same exact outcome from two consecutive measurements is impossible). 
Error correcting codes from the family of Slepian Wolf encoders are typically used in these systems; as an example, if the implementation is based on linear block codes, the helper information is typically in the form of the syndrome of the initial measurement.

From an information theoretic point of view, the basic idea is to generate a hashed version of the initial measurement, similarly to regenerating a unique secret key used for authentication, derived during the enrollment stage. As the rate of the secret key generation increases, the attacker has a harder task to guess it correctly. In other words, the level of security increases, while a lower bound on the length of the authentication key is imposed by the size of the brute force attack that can be mounted by an adversary.
In following, we describe three different PLS approaches employed for node authentication. We note that a combination of these can also be employed, in order to increase the authentication vector size.

\subsection{Physical unclonable functions (PUFs)}
The concept a physical unclonable function was first introduced in \cite{Pappu2002}. The idea is that integrated circuits (ICs) have uniquenesses in their physical microstructure which is inherited from  inevitable variations during the fabrication process. These unique characteristics are \textit{unpredictable} before the end of the manufacturing and can be considered as digital signatures or identities of the ICs. A PUF should be \textit{unclonable}, which means that given the exact fabrication procedure, it is infeasible to reproduce the same physical microstructure. The security of PUFs stems from these properties. An example of an arbiter based PUF is depicted in Fig. 2.

PUFs operate based on challenge-response pairs (CRPs). In the enrollment stage, a set of challenges are applied to a PUF (e.g., in the form of input voltages to a chain of logical gates) and the response of the PUF to each challenge, i.e., a hash of the PUF measurement, are stored in a database at the verifier. The responses for different challenges are different and each PUF has a unique response. Due to noise in measurements, in this stage some helper data is also stored in the database to enable the re-generation of the registered CRPs in the release stage. The collected sets of CRPs are considered as the IDs of the devices to be authenticated. 

In the release stage, the verifier presents a particular challenge to the PUF. After running the challenge, the PUF releases the corresponding response (PUF measurement). If the response collected from the PUF in the release stage along with the helper information can reproduce the stored authentication key in the enrollment stage, the user is authenticated. A PUF that has an exponential number of CRPs is considered ``strong'' \cite{Ruhrmair10}, i.e., it has a higher entropy and is a better option for security purposes as opposed to weak PUFs with polynomial numbers of CRPs. In \cite{Maubach06}, the authors consider an information theoretic perspective on PUFs and derive the entropy in a particular type of PUFs based on their physical properties.

In the CRP scheme, the attacker may acquire a software model of the PUF by using information extracted from exchanged CRPs in the clear. Intrinsically, a PUF hides a “random” function, and learning such functions from input-output pairs falls within the context of machine learning (ML). The authors in \cite{Nguyen19} show their proposed PUF is secure against the strongest known classical and reliability-based ML attack.
Different PUF-based authentication protocols for wireless sensor networks have recently been proposed in the literature \cite{Ersi0,Ersi19B,Ersi19C,Chatterjee17,Mahalat18,Aman19}.

\subsection{Biometrics}
Biometric authentication is used for user (as opposed to device) authentication. The security of the method comes from the  uniqueness and consistency of biometric characteristics of each person. Similar to PUF based authentication, the fundamental scheme consists of two stages. In the enrollment stage the biometric characteristics of users are sampled and in plain form or through a transformation are stored in the database of the verifier. Due to noisy measurement and possible changes in biometric characteristics over time or damages, helper data is also stored in this stage. In the release stage, the verifier demands a new biometric sample from the user and if the new measurement with the assist of the helper information can reproduce the same data stored during the enrollment stage, the user is authenticated.

This method of authentication has been widely used over decades for different applications. However, privacy concerns pose a major challenge. The biometric characteristics of a human cannot be changed. If the stored data are compromised by attackers, they can be used to imitate legitimate users. Different approaches have been proposed to protect the stored data from such attacks. For example, in \cite{Sutcu07,Sutcu07B} a type of cryptographic primitive called secure sketch is considered. In this approach, a hash of the biometric information is stored in the database along with the helper data. In \cite{Ratha07,Bringer08}, the authors study a cancelable biometric scheme in which an irreversible transformation of the biometric data is stored. 

Information theoretic analyses of these schemes have been performed \cite{Ignatenko07,Cohen04} and the largest rate of the authentication key has been characterized \cite{Tuyls04} in absence of privacy requirements. In all of the aforementioned works and in the basic proposed scheme, the helper data can contain information about the biometric characteristics. The two part paper \cite{Lai11A, Lai11B} studies the privacy-security trade off in biometric security and considers two scenarios with perfect key protection and perfect privacy model, that address two different perspectives of the problem.
\paragraph{Perfect key protection system}
In this model the helper data ($V$) does not contain any information about the secret key ($K$). The privacy of the biometric measurement is measured as the normalized equivocation rate $H(X^n|V)/H(X^n)$, where $H(\cdot)$ denotes entropy. The greater normalized equivocation means the higher level of privacy which can be arbitrarily close to unity when the mutual information of $V$ and $X$ goes to zero ($I(V;X)\to 0$). In perfect key protection system, there is a trade off between the rate of secret key generation $R$ and the level of the biometric measurement privacy $\Delta_P$. For a perfect key protection biometric authentication system, a privacy-security pair $(\Delta_P,R)$ is said to be achievable if for any $\epsilon>0$, there exists an integer $n$, that satisfies the following conditions:
\begin{align}
H(K)/n &\geq R, \\ 
H(X^n|V)/H(X^n)&\geq \Delta_P, \\
I(V;K)/n &\leq \epsilon, \\
Pr(K \neq \hat{K})&\leq \epsilon,
\label{PerfectKeyConst}
\end{align}
where $X$ and $\hat{K}$ represent the measurement in the enrollment stage and the estimated secret key, respectively. It has been shown that the capacity region $\mathbb{C}$ contains the set of all privacy-security pairs $(\Delta_P,R)$ such that \cite{Lai11A}
\begin{align}
\Delta_P &\leq 1 - \frac{I(U;X)-I(U;Y)}{H(X)},\\ 
R&\leq I(U;Y),
\label{PerfectKeyCapacity}
\end{align}
where $Y$ denotes the measurement in the release stage and $U$ is an auxiliary random variable such that $(U,X,Y)$ forms a Markov chain $U \to X \to Y$.

\paragraph{Perfect privacy system}
In this model the helper data ($V$) does not contain any information about the biometric measurement ($X$). The performance of perfect privacy system can be measured by the rate of secret key generation and the normalized equivocation of the generated key $H(K|V)/H(K)$. In this system, a rate-equivocation $(R,\Delta_s)$ is achievable if for any $\epsilon>0$, there exists an integer $n$, that satisfies the following conditions:
\begin{align}
H(K)/n &\geq R, \\
I(X^n;K)/n &\leq \epsilon, \\
H(K|V)/H(K)&\geq \Delta_s, \\
Pr(K \neq \hat{K})&\leq \epsilon.
\label{PerfectPrivacyConst}
\end{align}
It has been shown that a privacy-rate pair $(R,\Delta_s)$ is achievable if and only if, for the random processes $X^n$ and $Y^n$ for each $\epsilon > 0$ there exist an $n$ and functions $\Psi_n$ of $X^n$ and $\Phi_n$ of $Y^n$ such that \cite{Lai11A}
\begin{align}
Pr [\Psi_n(X^n)\neq \Phi_n(Y^n)]&\leq \epsilon , \\
H(\Psi_n(X^n))/n &\geq R\Delta_s-\epsilon.
\label{PerfectPrivacyConst}
\end{align}
Note that if $K$ is a function of $X^n$, perfect privacy means perfect key protection.
\subsection{Wireless identification using RF fingerprinting}
Utilizing wireless channel characteristics is another approach to authenticate the nodes. In this approach, the wireless channel characteristics such as the user / device localization, e.g., using the received signal strength indicator (RSSI) or the link quality indicator (LQI) and the angle of arrival are used to verify the ``expected location'' of the users / devices. Wireless identification is commonly used in scenarios in which localization also needs to be verified e.g., in IoT sensors monitoring temperature and pressure at various equipment. Different variants of wireless identification have been studied for different applications \cite{Aman19,Rasmussen07,Aman17,Ali13}.

\section{Message Integrity}\label{MssgIntg}
A second major requirement of secure communications is that the legitimate receiver should be able to ensure the integrity of received messages. In many applications, this operation is considered even more important than that of confidentiality, given that many messages might not be ``secret'' but should be ``authentic''. 
In this scenario, the opponent is active and can sketch different attacks to deceive the receiver, typically by tampering with the message in transit. As an example, in substitution attacks, the attacker changes content of the message transmitted by the legitimate source. In impersonation attacks, the attacker sends a fake message while the source is idle. The receiver should be able to detect the fake and modified messages from the authentic ones \cite{Ersi13C,Ersi12}.

Message integrity requires a secret key shared by the two communicating parties and unknown by the attacker. The rest of the system design,  such as the encoding / decoding schemes, are publicly available. The receiver considers the received signal as authenticated / verified (i.e., the integrity test is successful), if there exists a valid ``tag'' that can uniquely relate the received message to the secret key $k$, while the attacker cannot produce a valid (tag, message) pair despite intercepting a large number of related exchanges; proofs along this line of reasoning fall into the category of chosen ciphertext semantic security. 
The entropy of the shared key should be high enough to not allow the attacker to mount brute force attacks on the system. 

\begin{figure}[t]
\centering
\includegraphics[width=0.48\textwidth]{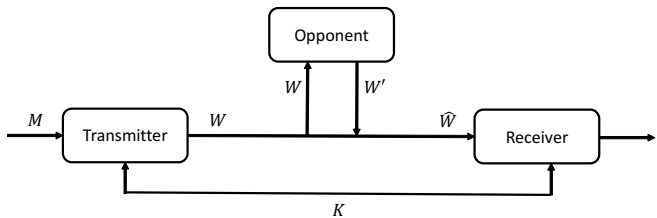}
\caption{The message authentication model for noiseless channel.}
\label{mssg1}
\end{figure}

The information theoretic limits of the message authentication problem was first considered by Simmons \cite{Simmons85}. In Simmons's model, a noiseless channel between the terminals has been assumed (Fig. \ref{mssg1}). In this model, the transmitter transmits $w$ which is a function of the secret key $k$ and the message $m$ ($w=f(k,m)$). The opponent observes the signal  transmitted in the clear. However, the receiver observes $\hat{w}$, which can be different from the original signal and modified by the opponent. It has been shown that the success probability of impersonation and substitution attacks in message authentication can be lower bounded by $2^{-I(K,W)}$ and $2^{-H(K|W)}$, respectively. Therefore the success probability of the attacks by the opponent is lower bounded by $\frac{1}{\sqrt{|\mathcal{K}|}}$, where $|\mathcal{K}|$ is the size of the key space.

\begin{figure}[t]
\centering
\includegraphics[width=0.48\textwidth]{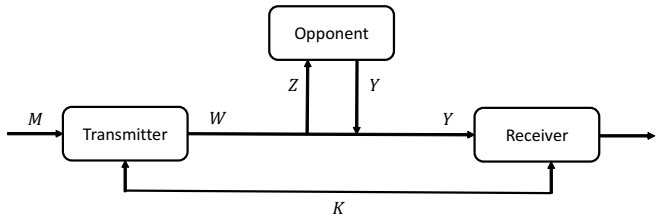}
\caption{The message authentication model for noisy channel.}
\label{mssg2}
\end{figure}

In \cite{Liu05, Boncelet06}, the authors consider noisy channels and demonstrate that introducing noise in the model can make the receiver reject some valid messages. They conclude that channel noise is detrimental to message authentication. Coversely, in \cite{Lai09, Lai10}, the authors study the message authentication problem via noisy channels from a new perspective. The authors propose a scheme in which the transmitter exploits the noise in the channel to ``hide'' the key information from the opponent. In their scheme, the transmitter performs joint channel coding and message authentication coding. The channel code is designed such that the conditional probability distribution after observing the channel output at the opponent side is very close to a uniform distribution. In \cite{Lai09}, a discrete memoryless channel (DMC) model is considered and it is assumed that the opponent observes $Z$ with a particular conditional probability distribution given the message $w$ has been sent (Fig. \ref{mssg2}). If the opponent does not perform any attack, the receiver observes $Y$ with a particular conditional probability distribution given $w$. However, if the opponent performs an attack, it can modify $Y$ according to its attack policy.

The receiver may make two possible types of error, which are to wrongly reject an authentic message (false negative)) or to accept a modified or fake message (false positive). The proposed scheme in \cite{Lai09} utilizes a wiretap channel model to protect the secret key. Basically, the transmitter chooses an input distribution $P_W$ such that $I(W;Y)-I(W;Z)>0$. Then, the source generates a codebook for the wiretap channel with $2^{nI(X;Y)}$ codewords, where $n$ is the blocklength and we assume it is large enough to satisfy a low decoding error probability requirement at the receiver. The source then partitions the codewords into $|\mathcal{K}|$ subsets, i.e., $|\mathcal{K}|<2^{n[I(W,Y)-I(W,Z)]}$. Each subset is associated  with each key. 

Assume, the codeword length be large enough that there be more than $|M|$ codewords in each subset. The source then divides each subset into $|M|$ bins, each corresponding to a message. There are multiple codewords in each bin. In the transmission, if the intended message is $m$, and the key is $k$, the source then randomly chooses a codeword $w$ from the $m$th bin of the $k$th subset using a uniform distribution. As the coding rate is $I(X;Y)$ the receiver can decode the message with high probability if $n$ is large enough. On the other hand, according to the fundamental wiretap channel result, the opponent cannot gather a significant amount of information about the secret key in this scheme. It is shown that the success cheating probability is upper bonded by $\frac{1}{|\mathcal{K}|}$, which is significantly higher than the bound for the noiseless channel;  as this scheme the transmitter uses the noise of the channel an an added entropy source to ``hide'' the key, the same approach cannot be applied for the noiseless scenario.

\section{Message Confidentiality Using Secrecy Encoders}\label{MssgConf}
Next, we study two distinct approaches to keep the messages confidential from third parties: i) the wiretap channel model, which exploits an advantage in terms of channel quality at the legitimate receiver in this Section,  and, ii) secret key generation from shared randomness which exploits a common alea shared by the legitimate pair and at least partially unobserved by the opponent to generate a secret key (e.g., to be used with some appropriate encryption scheme) \cite{Ersi16A}, in the next Section.

Wyner in \cite{Wyner75} introduced the discrete memoryless wiretap channel model. In this model the transmitter communicates with a legitimate receiver and they want to keep the message confidential from a third party who is eavesdropping  (passively intercepting the channel as in Fig. \ref{wiretap_fig}). In this scheme, no secret key is shared between the legitimate nodes. Wyner showed that the maximum achievable rate at which both reliable communication between the legitimate parties and weak secrecy with respect to the eavesdropper can be established, referred to as the channel's secrecy capacity $C_s$, can be expressed as follows:
\begin{figure}[t]
\centering
\includegraphics[width=0.48\textwidth]{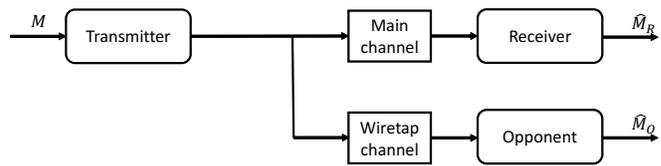}
\caption{The wiretap channel.}
\label{wiretap_fig}
\end{figure}
\begin{align}
C_s = \max_{V\to X\to YZ} I(U;Y)-I(U;Z),
\label{Wiretap_Eq}
\end{align}
where $Y$ is the observation of the legitimate receiver, $Z$ denotes the observation of the eavesdropper, $X$ is the transmitted codeword and $U$ is an auxiliary random variable such that $(U,X,YZ)$ is a Markov chain $U\to X\to YZ$. According to (\ref{Wiretap_Eq}), the secrecy capacity is the difference maximization between two values of mutual information which is taken over all possible input distributions $p(x)$. Hence both the legitimate receiver and the opponent channel conditions are essential to wiretap code designs. As it is mentioned before, in this model, the legitimate nodes do not need to share any secret key for their communication. One of the main drawbacks of using wiretap channel encoders in practice is that in this model the secrecy of the communication can only be established when there exists a particular input distribution $p(x)$ such that the mutual information of the main channel is higher than that of the wiretap channel, which is not guaranteed. Importantly, the transmitter needs to know the channel state of the opponent, an assumption that is impractical in many scenarios.

As a solution to the latter concern, the wiretap channel model with partial channel state information have been studied \cite{Mukherjee11,Huang12}, using an appropriate model where the uncertainty of practical CSI is taken into account. In the related model, the wiretap channel coefficient is divided to two parts. The first part is assumed to be known by the transmitter while the second one is unknown. As the weight of the second part becomes higher, the transmitter has lower information of the wiretap channel coefficient. In the case that the wiretap channel state is not available, the secrecy outage probability of is alternatively used as the security performance metric, in lieu of the secrecy capacity. The secrecy outage probability indicates the probability that the instantaneous secrecy capacity $C'_s$ is lower than a target value $C_s$. 

Furthermore, a plethora of alternative techniques have also been proposed in the literature to mitigate the need for full adversarial CSI, such as transmitting in the adversary's null signal space by leveraging the potential of multiple-input multiple-output (MIMO) transmission, injecting artificial noise to the adversarial signal space \cite{Ersi12B,Ersi12C}, adaptive power allocation \cite{Ersi15B, Ersi14B,Ersi13B}, exploitation of relay channels, faster than Nyquist assisted secrecy \cite{Ersi11,Ersi10}, network coding \cite{Ersi16B, Ersi14}, and cognitive radio systems \cite{Poor17,Chen17}.

\section{Secret Key Generation (SKG) from Wireless Fading Coefficients}
\begin{figure}[t]
\centering
\includegraphics[width=0.48\textwidth]{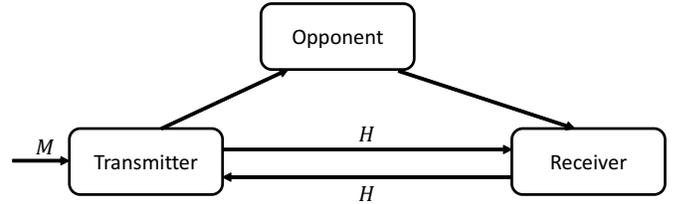}
\caption{The channel based secret key generation system model.}
\label{SKG}
\end{figure}

In this Section, we review the generation of secret keys from common randomness in the form of the wireless channel coefficient observed by a transmitter / receiver pair. This approach exploits the reciprocity of the wireless channel during the channel coherence time. The reciprocity refers to the property that the channel responses at both sides are the same  (Fig \ref{SKG}). Therefore, the two endpoints of the channel observe a noisy form of a  shared randomness from which they can distil a secret key. In a multipath rich environment, a third party should be only a few wavelengths away from the transmitter or the receiver so that their observed channel coefficients are independent from that in the direct channel between the transmitter and the receiver; this is a practical assumption in many actual wireless scenarios. Such generated keys can be secure with information-theoretic guarantees when the generation scheme is carefully applied, as opposed to keys generated from pseudorandom number generators.

The SKG standard procedure typically encompasses three phases \cite{Maurer93}:
\paragraph*{Advantage distillation} the legitimate nodes exchange probe signals to obtain estimates of their reciprocal channel state information (CSI) and pass them through a suitable quantizer \cite{Wang11}. Commonly, the received signal strength (RSS) has been used as the CSI parameter for generating the shared key \cite{Mathur08}, while in \cite{Ersi15, Sayeed08} the CSI phase has been proposed.

\paragraph*{Information reconciliation} discrepancies in the quantizer local outputs due to imperfect channel estimation are reconciled through public discussion using Slepian Wolf decoders. In this phase, with the aid of public discussion, the nodes should reconcile to a common key while avoiding to reveal any information about it. Numerous practical information reconciliation approaches using standard forward error correction (FEC) codes such as low density parity check codes (LDPC) have been proposed \cite{Ye06, Ye10}, while in \cite{Ersi15} the possibility of employing short Bose, Chaudhuri, Hocquenghem (BCH) FEC codes has also been explored.

\paragraph*{Privacy amplification} applying universal hash functions to the reconciled information ensures that the generated keys are uniformly distributed (i.e., have maximum entropy) and are completely unpredictable by an adversary \cite{Maurer07}. More importantly, it ensures that even if an adversary has access to (even a large) part of the decoder output, the final secret key can be unpredictable \cite{Bloch11}. However, in this case, the genuinely random input space to the hash function needs to be large enough in order to avoid brute force attacks. When part of the reconciled information is known to the adversary, the corresponding amount of entropy needs to be ``supressed'' by the privacy amplifier.

Employing the standard SKG system model, let us assume that the transmitter and the receiver exchange a probe signal $X$ in two consecutive slots and that their respective observations $Z_A$ and $Z_B$, can be expressed as
\begin{align}
Z_A &= X H + N_A, \\
Z_B &= X H + N_B,
\end{align}
where $X$ denotes the channel input and $H$ is the channel gain between the legitimate nodes, modeled as a circularly symmetric complex Gaussian random variable with zero mean and variance $\sigma_H ^2$. $N_A$ and $N_B$ denote accordingly circularly symmetric complex Gaussian zero mean random variables that model the impact of additive white Gaussian noise with variances $\sigma_A ^2$ and $\sigma_B ^2$, respectively (typically $\sigma_A ^2=\sigma_B ^2$). 

\subsection{Secret key rate}
At first, let us assume that the attacker is a passive eavesdropper that only tries to obtain an estimate of $H$ by interception. The case of an active attacker will be considered next. The information theoretic limits regarding the rate for generating secret keys has been established in  \cite{Ahlswede93}. From an information theoretic perspective, a secret key with rate $R$ is achievable if for any $\epsilon > 0$ and sufficiently large blocklength $n$, there exists a public discussion strategy such that 

\begin{align}\label{SKG_def}
Pr{(K_A\neq K_B)} &<\epsilon ,\\
\frac{1}{n}I(\Phi,\Psi;K_1) &<\epsilon , \\
\frac{1}{n} H(K_1) &>R - \epsilon , \\
\frac{1}{n} \log(|\mathcal{K}|)&<\frac{1}{n}H(K)+\epsilon,
\end{align}
where $\Phi$ and $\Psi$ denote the public messages sent by the transmitter and receiver in the information reconciliation sub-process, respectively, and $K_A$ and $K_B$ denote the distilled keys by the transmitter and receiver, respectively.

\begin{theorem}
The secret key capacity $C_s$ assuming unlimited public discussion case is given as \cite{Maurer93}
\begin{align}
C_s = I(Z_A;Z_B).
\end{align}
\end{theorem}
However, in some scenario, there may be some limitations on public channel discussion. The capacity of secret key in public channel with limited rate is discussed in the following theorem.

\begin{theorem}
The secret key capacity $C_s$ when the public channel rate constraint is $R$, is given by \cite{Csiszar00}
\begin{align}
C_s = &\max_{U} I(U;Z_B), \\
s.t. \quad &U\to Z_A \to Z_B , \\
&I(U;Z_A)-I(U;Z_B)\leq R,
\end{align}
where $U$ is an auxiliary random variable.
\end{theorem}

Furthermore, it is possible that the evesdropper observes a sequence $Z_E$ correlated to the the common randomness source. In this case, the security constraint in (\ref{SKG_def}) should be transformed to
\begin{align}
\frac{1}{n}I(\Phi,\Psi,Z_E;K_1) <\epsilon.
\end{align}
The secret key capacity $C_s$ when the opponent has side information $Z_E$ and the public channel rate constraint is $R$, is in the following theorem.

\begin{theorem}
The secret key rate $R_s$ is achievable when the opponent has side information $Z_E$ and the public channel rate constraint is $R$, is \cite{Csiszar00}
\begin{align}
R_s = &[I(U;Z_B)-I(U;Z_E)]^+ , \\
s.t. \quad &U\to Z_A \to Z_B, \\
&I(U;Z_A)-I(U;Z_B)\leq R,
\end{align}
where $U$ is an auxiliary random variable and $[x]^+=\max\{x,0\}$.
\end{theorem}

\subsection{Authenticated encryption using SKG} \label{subsec:key_integrity}

Under the system model in Fig. \ref{SKG} and normalizing to unity the noise variances, i.e., $\sigma_A ^2=\sigma_B ^2=1$) for simplicity, the SKG rate can expressed as \cite{ Ersi17,Ersi17B,Ersi17C}:
\begin{equation}
R_k=\log_2 \left(1+ \frac{P \sigma^2}{2+\frac{1}{P \sigma^2}} \right), \label{eq:SKG_rate}
\end{equation}
while the corresponding \textit{minimum} necessary reconciliation rate has been shown to be $h(H_{B}|H_{A})$ \cite{Ahlswede93}, where $h(\cdot)$ denotes differential entropy.
To develop a hybrid cryptosystem that can withstand active attacks \cite{Ersi17D,Ersi16C}, the SKG can be used in conjunction with standard block ciphers, e.g., AES in Galois counter mode (GCM), to build hybrid authenticated encryption schemes. 

As a sketch of such a hybrid scheme, let us assume a system with three parties: Alice who wishes to transmit a secret message $\mathbf{m}$  to Bob with confidentiality and integrity, and Eve (the opponent), that can act as a passive and active attacker. The following algorithms are employed: 
\begin{itemize}
    \item The SKG scheme denoted by $\verb"G": \mathcal{H}\rightarrow \mathcal{K} \times \mathcal{S}$, accepting as inputs a vector of complex numbers (the fading coefficients),  and generating as output a binary vectors of sizes $n$ and $n-k$, respectively, $n,k \in \mathbb{N}$, (in the key and the syndrome spaces), \textit{i.e.},
    \begin{equation}
        \verb"G"(H)= \left(K, S_{A}\right),
    \end{equation}
  where $K\in\mathcal{K} $ denotes the key obtained from $H$ after privacy amplification and $S_A \in \mathcal{S}$ is Alice's syndrome (side information used for reconciliation).
\item A symmetric encryption algorithm, e.g., AES GCM, denoted by $\verb"Es": \mathcal{K}\times \mathcal{M} \rightarrow \mathcal{C} $ where $\mathcal{C}$ denotes the ciphertext space with corresponding decryption $\verb"Ds": \mathcal{K}\times \mathcal{C} \rightarrow \mathcal{M}$, such that 
\begin{eqnarray}
    \verb"Es"(K,m)=c,\\
    \verb"Ds"(K, c)=m,
\end{eqnarray}
for $K\in \mathcal{K}$, $m\in \mathcal{M}$, $c\in\mathcal{C}$.

\item A pair of message authentication code (MAC) algorithms, e.g., in HMAC mode, denoted by $\verb"Sign": \mathcal{K}\times \mathcal{M}\rightarrow \mathcal{T}$, with a corresponding verification algorithm $\verb"Ver": \mathcal{K}\times \mathcal{M} \times \mathcal{T} \rightarrow (yes, no)$, such that 
\begin{eqnarray}
&&\verb"Sign" (K, m)=t,\\
&&\verb"Ver" (K, m, t)=\left\{\begin{array}{ll}
\it{yes}, & \text{if integrity verified}\\
\it{no}, & \text{if integrity failed}
\end{array}\right. 
\end{eqnarray}
\end{itemize}

A hybrid crypto-PLS system for AE SKG can be built as follows:
\begin{enumerate}
    \item The SKG procedure is launched between Alice and Bob generating a key and a syndrome $\verb"G"(H)\!=\!(K,S_A)$. 
    \item Alice breaks her key into two parts $K=\{K_e, K_i\}$  and uses the first to encrypt the message as  $c=\verb"Es"(K_e, m)$. Subsequently, using the second part of the key she signs the ciphertext using the signing algorithm $t=\verb"Sign"(K_i, c)$ and transmits to Bob the extended ciphertext $\left[S_A\| c\| t\right]$, where $\left[\cdot\| \cdot\right]$ denotes concatenation of the corresponding binary vectors. %
    
    \item Bob checks first the integrity of the received ciphertext as follows: from $\mathbf{S}_A$ and his own observation he evaluates $K=\{K_e, K_i\}$ and computes $\verb"Ver"(K_i, c, t)$. The integrity test will fail if any part of the extended ciphertext was modified, including the syndrome (that is sent as plaintext); for example, if the syndrome was modified during the transmission, then Bob would not have evaluated the correct key and the integrity test would have failed. 
    \item If the integrity test is successful then Bob decrypts $m\verb"=Ds"(K_e, c)$.
\end{enumerate} 

\subsection{Shielding SKG from active attacks during pilot exchange}

The proposed authenticated encryption scheme using SKG is however vulnerable to man-in-the-middle (MiM) attacks during the pilot exchange phase, a vulnerability that was until recently unexplored. Here, we propose a simple scheme to overcome this issue. 
 We assume a man-in-the-middle (MiM) attack in the form of an injection signal and then move to denial of service attacks in the form of jamming \cite{Ersi18,Ersi19A,Ersi13}.

\paragraph*{MiM attacks} MiM in SKG pilot exchange takes the form of a signal injection. Various possible approaches have so far surfaced on how to launch injection attacks; the attack can consist in controlling the movement of intermediate objects in the wireless medium, thus generating predictable changes in the received RSSI (e.g., by obstructing or not the line of sight), or the opponent can spoof the SKG process by injecting a signal $W$ to Alice and Bob so the opponent will have some information about the secret key, as follows
\begin{align}
Z_A=XH+W+N_A, \nonumber\\
Z_B=XH+W+N_B, \label{eq:Zb}
\end{align}
where $W$ denotes the injected signal. A simple approach to generate $W$ can be devised as long as the adversary has one more antenna than the legitimate users. An example, let us consider the case in which Alice and Bob have one antenna each and the MiM has two. In this case, the MiM can choose a precoding matrix $P$ so that

\begin{align} 
W=\mathbf{H_{AE}}^T\mathbf{P} X_J=\mathbf{H_{BE}}^T\mathbf{P}X_J
\end{align}
where, $\mathbf{H_{AE}}$ and $\mathbf{H_{BE}}$ denote the channel matrices between Alice and Eve and Bob and Eve, respectively. 
The precoding matrix $\mathbf{P}$ can be built as follows:
\begin{align}
    \mathbf{H_{AE}}^T\mathbf{P} X_J=\mathbf{H_{BE}}^T\mathbf{P}X_J
    \Rightarrow  P_1=\frac{H_{BE2}-H_{AE2}}{H_{AE1}-H_{BE1}}P_2,
\end{align}
where $X_J$ is a generic transmitted signal by the MiM to satisfy the power constraint. 

Under this attack, the secret key rate controlled by the opponent is upper bounded by \cite{Ersi18}
\begin{equation}
L \leq I(Z_A, Z_B; W).
\end{equation}
A countermeasure to injection attacks can be built by randomizing the pilot sequence exchanged between Alice and Bob  \cite{Ersi18}. Here, we propose to randomize the pilots by drawing them from a (scaled) QPSK modulation, as follows: 
instead of transmitting the same probing signal $X$, Alice and Bob transmit independent, random QPSK probe signals $X$ and $ Y$, respectively. Alice's  observation $Z_A$ is modified accordingly as 
\begin{eqnarray}
{Z}_A&=&YH+W+N_A,
\end{eqnarray}
while Bob's observation is given in (\ref{eq:Zb}).
To establish shared randomness in spite of the pilot randomization, Alice and Bob post-multiply $Z_A$ and $Z_B$ by their randomized pilots, obtaining local observations $\tilde{Z}_A$ and $\tilde{Z}_B$ (unobservable by Mallory), expressed as:
\begin{eqnarray}
\tilde{Z}_A&=&X Z_A=XYH+XW+XN_A,\\
\tilde{Z}_B&=&YZ_B=XYH+YW+YN_B.
\end{eqnarray}
The source of shared randomness, when the pilots are randomized QPSK symbols, is a circularly symmetric zero mean Gaussian random variable, $XYH \sim C\mathcal{N}(0, P^2\sigma^2)$.

Furthermore, due to the fact that $X$ and $Y$ are independent and have zero mean, the variables $XW$ and $YW$ are uncorrelated, circularly symmetric zero-mean Gaussian random variables, and, therefore independent, while the same holds for $XN_A, YN_B$, i.e., $(XW, YW) \sim \mathcal{CN}(\mathbf{0}, \sigma_J^2P\Gamma\mathbf{I}_2)$ and $(XN_A, YN_B) \sim \mathcal{CN}(\mathbf{0}, P\mathbf{I}_2)$. Alice and Bob extract the common key from the modified source of common randomness $XYH$ as opposed to $XH$. On the other hand, since $XW, YW, X N_A, Y N_B$ are i.i.d. complex circularly symmetric Gaussian random variables, the proposed scheme reduces injection attacks to uncorrelated jamming attacks, i.e., we get that \begin{equation}
    L\leq I\left(\tilde{Z}_A, \tilde{Z}_B; W   \right)=0.
\end{equation}
Pilot randomization has in essence reduced injection attacks to jamming attacks.
\paragraph*{Jammming attacks} Building on the results of the previous subsection, we next examine in detail the scenario in which the attacker acts as a jammer. 
Tow major alternatives have been identified to counter jamming:

\subsubsection{The legitimate nodes harvest energy (EH) from the jamming signal}
By harvesting the jamming power in a first phase and exploiting it to boost the pilot power during SKG in a second phase, the jammer’s action may in fact increase the SKG capacity; in this case, the jammer should not launch the attack, i.e., it is neutralized. However, it is not always optimal for the legitimate nodes to neutralize the jammer. Indeed, using EH can reduce the SKG capacity since, for a non-trivial fraction of time, there is no secret bits generation; when the jammer is neutralized the penalty in terms of SKG rate might become too high, depending on the system parameters [reference missing].
\subsubsection{Channel hopping  or spreading}
If the legitimate nodes do not have EH capabilities, yet there is another way to defend against jamming by assuming that the legitimate nodes can employ channel hopping or spreading over multiple orthogonal subcarriers [paper reference missing]. 

Here, the idea is to use channel hopping in a random fashion and avoid most of the jammer’s interference as opposed to completely neutralizing it. Since potential jammers cannot predict the subcarrier used by the legitimate nodes, they will always spread their powers over the entire spectrum: the larger the number of subcarriers, the smaller the jammer’s interference on each subcarrier. However, channel hopping is not always optimal since only a fraction of the entire spectrum is used for SKG. Depending on the system parameters, it can be preferable for the legitimate nodes to spread the available power across the entire spectrum rather than concentrate it on a single subcarrier.

\section{Conclusion}\label{Conc}
Many standard cryptographic schemes, particularly those in the realm of public key encryption (PKE), are computationally intensive, incurring considerable overhead. For example, a 3GPP report on the security of ultra reliable low latency communication (URLLC) systems notes that ``for a URLLC service with higher speed than 65 kbps, the 3GPP Release 15 radio access network (RAN) cannot fulfill the quality of service (QoS) requirement while enforcing user plane integrity protection"~\cite{3gppURLLC}.  Additionally, traditional public key generation schemes are not \textit{quantum secure} -- in that when sufficiently capable quantum computers will be available they will be able to break current known public key encryption schemes -- unless the key sizes increase to impractical lengths. 

In this chapter, we have reviewed alternative approaches to secure future communication systems by considering PLS. We have presented recent results on PLS with respect to major emerging application areas in authentication, integrity and confidentiality. We have focused on three topics of secure communications, namely node authentication, message integrity, and, secrecy, including secret key generation. We have reviewed some of the information theoretic limits and discussed implementations proposed recently in the literature. Additionally, we have discussed open issues that need to be addressed before the employment of PLS in future generation networks.

\section{Acknowledgments}\label{Ack}
This work is supported by the ELIOT ANR-18-CE40-0030.
\bibliographystyle{IEEEtran}
\bibliography{bibliography}

\end{document}